\documentstyle[psfig]{l-aa}

\newcommand{\tablehead}[1]{\gdef\tableDhead{#1}}
\newcommand{\tabletail}[1]{\gdef\tableDtail{#1}}
\newenvironment{MonTabular}[1]{
  \begin{tabular}{#1}
  \tableDhead
}{
  \tableDtail
  \end{tabular}
}

\newcommand{\mm}[1]{\ifmmode #1\else$#1$\fi}

\newcommand{\Ly}[1]{\mm{\mathrm{Ly}#1}}
\newcommand{\m}[1]{\mm{m_{\mathrm{#1}}}}
\newcommand{\M}[1]{\mm{M_{\mathrm{#1}}}}
\newcommand{\z}[1]{\mm{z_{\mathrm{#1}}}}

\newcommand{\Lya}[0]{\Ly{\alpha}}
\newcommand{\Lyb}[0]{\Ly{\beta}}
\newcommand{\forb}[2]{[{\rm #1}\,{\sc #2}]}
\newcommand{\MgII}[0]{{\rm Mg}\,{\sc ii}}
\newcommand{\FeII}[0]{{\rm Fe}\,{\sc ii}}
\newcommand{\HI}[0]{{\rm H}\,{\sc i}}
\newcommand{\HII}[0]{{\rm H}\,{\sc ii}}
\newcommand{\OII}[0]{\forb O{ii}}
\newcommand{\NII}[0]{\forb N{ii}}
\newcommand{\OIII}[0]{\forb O{iii}}
\newcommand{\Wr}[0]{\mm{w_{\mathrm{r}}}}
\newcommand{\mr}[0]{\m{r}}
\newcommand{\mb}[0]{\m{B}}
\newcommand{\Mr}[0]{\M{r}}

\newcommand{\Mab}[0]{\M{AB}}
\newcommand{\ze}[0]{\z{e}}
\newcommand{\za}[0]{\z{a}}
\newcommand{\zg}[0]{\z{g}}
\newcommand{\kms}[0]{\mm{\mathrm{km~s}^{-1}}}
\def\h50{\mm{h_{50}^{-1}}}

\begin{document}

\thesaurus{03(03.12.1; 07.07.1; 07.10.1; 17.01.1) }
 
\title{Evolution of quasar absorption-selected galaxies
\thanks{Based on observations made at the European Southern Observatory, La
Silla, Chile (programme 1-012-43K)}}

\author{P. Guillemin \inst{1}
  \and J. Bergeron \inst{2,1}}
\offprints{J. Bergeron (jbergero@eso.org)}
\institute{
  Institut d'Astrophysique de Paris, CNRS, 98bis Boulevard Arago, F-75014 
  Paris, France.
  \and
  European Southern Observatory, Karl-Schwarzschild Stra$\ss$e 2, D-85748 
  Garching bei M\"unchen, Germany.
}

\date{received 19 August 1996 ; accepted 21 May 1997}

\maketitle
 
\begin{abstract}
We present the results of a survey of galaxies selected by their gas
cross-section which give rise to \MgII\ absorption lines in the
spectra of background quasars. The sample comprises 11 galaxies  covering
the redshift range $0.7<z<1.2$ with $\left<z\right>$=0.9, and  is
combined with  a lower redshift ($z<0.7$) sample of 15 \MgII\ absorbing galaxies.
 We first describe the properties of these two samples of galaxies whose
spectra range from present-day elliptical to irregular galaxies with a
dominance of Sbc and Scd types. The strong correlation found previously at
lower redshift between the halo radius and the galaxy luminosity still holds
at  $z\simeq1$, which implies no significant evolution in redshift of the
gaseous halo sizes. We find no evidence of a correlation between the
rest-frame equivalent width of the \OII$\lambda 3727$ emission line and the
rest-frame luminosity \Mab(B). This could be a consequence of the very blue
B$-$K color of intrinsically faint absorbing galaxies in the infrared, thus
optically bright as found in other surveys.  For the low and high redshift 
($z\leogr 0.7$) samples, we have built
template spectra by combining individual galaxy spectra.
The lack of detection of the  \NII\
emission line in the low-redshift template, \NII$\lambda$6568/H$\alpha <$0.2,
together with the values obtained for empirical abundance indicators,
suggests a O/H abundance in the disk of \MgII\ absorbing galaxies
 in the range [O/H]=$-0.6,-0.1$.
The spectral energy distribution (SED) of these templates at
$\lambda_{\mathrm{r}} >$3646\AA\ can be fitted by present-day Sbc or Sc
galaxies, but the UV excess at $\lambda_{\mathrm{r}} <$3500\AA\ follows the
SED of Scd and Im present-day galaxies at $\left<z\right>$=0.4 and 0.9
respectively. This blueing of the SED with redshift is associated with an increase of 
\Wr(\OII$\lambda 3727$) by about 45\%. 
To investigate the past history of star formation
of the \MgII\ absorbing galaxies, we have used the spectral evolution models
of stellar populations developed by Bruzual \& Charlot as well as red and blue 
absorber subsamples (individual galaxies with a Ba + Ca\,{\sc ii} 
break $\leogr 1.55$). The red subtemplates 
are well fitted by a Sb galaxy and the derived ages, 4.5 and 8.3 Gyr at
$z$=0.9 and 0.4 respectively, implies  a similar galaxy formation epoch
for the $\left<z\right>$=0.4 and 0.9 \MgII\ absorbing galaxies.
Results for the blue subtemplates using  Sc galaxy models, especially
that at   $\left<z\right>$=0.9, suggests that intense star formation activity
is occuring at $z\simeq$1, as also found for larger field galaxy samples at similar
redshifts.

\end{abstract}
 

\section{Introduction }

Several approaches have been used to study the nature and evolution of field
galaxies up to redshifts $z\sim 1$. One is to obtain redshift data on large
samples of galaxies, selected from their apparent magnitude in various
wavelength bands. These surveys now benefit from the multiplexing gain of
multi-slit spectrographs and several samples, each of few hundreds of
galaxies have recently been analysed. Another approach is to select galaxies
which produce metal absorption lines in the spectra of background quasars. 
The selection criterion is then the cross-section of the extended gaseous
component. The samples of \MgII\ absorption-selected galaxies are
fairly small since there are usually only one or two absorbers per field
and the number of known \MgII\ absorption systems is limited (the
two largest \MgII\ absorption surveys comprise 112 systems in the
redshift interval 0.2-1.5: Sargent et al. 1988a and 1988b; Steidel \& Sargent
1992).

The field galaxy samples selected on the basis of apparent magnitude are well
suited to derive the evolution of the galaxy number density and luminosity
function. Despite the excess counts of faint galaxies to \mb=24, the
$\mb<22.5$ selected samples at $\left<z\right>\sim 0.2$ and 0.3 of Broadhurst
et al. (1988) show a N($z$) distribution as that predicted for no evolution
model. This suggests an evolution in the apparent number density of galaxies
and/or a differential luminosity evolution of $L\la L^\star(z=0)$ galaxies
(Ellis 1993; Glazebrook et al. 1995). The analysis of a K-band-selected
galaxy sample by Songaila et al. (1994) suggests that there is no
significant, positive luminosity evolution out to $z$=1, and that
the brightest  galaxies may have even faded by  $z$=1. From the study of the Canada 
France Redshift Survey (CFRS), a I-band-selected sample with redshift data for
$I\le 22.5$ objects, Lilly et al. (1995) conclude that there is no evolution
in either the number density or luminosity function of galaxies redder than a
present Sbc galaxy over the redshift range $0<z<1$. In contrast, they
find a strong evolution of the luminosity function of galaxies bluer than a
present Sbc galaxy at $z>0.5$. Imaging with {\it HST} shows that a large
fraction of the latter are irregular/peculiar galaxies and this population
should be responsible for the steepening of the number-magnitude relation at
faintest magnitudes (Glazebrook et al. 1995) and of the observed
differential evolution of the galaxy luminosity function (Schade et al.
1995).

Deep surveys of quasar absorption line-selected galaxies provide samples of
galaxies a-priori not biased against magnitude, color or morphology. The
first survey of \MgII\ absorption-selected galaxies by Bergeron \& 
Boiss\'e (1991) has shown that at $\left<z\right>\simeq0.4$, all field
galaxies brighter than $L\ga 0.3L^\star$ should have extended gaseous halos
with a typical radius $R(L^\star)\sim 75\h50$ kpc ($h_{50}$ is the Hubble
constant in units of 50 \kms Mpc$^{-1}$: we assume throughout this paper 
$h_{50}$=1  and $q_0$ = 0). The distribution 
of \OII$\lambda 3727$ rest-frame equivalent width is close to that observed 
for field galaxy samples at similar redshifts. The lack of
underluminous galaxies, $L<0.3L^\star$, with extended gaseous halos has been
confirmed by a larger redshift sample of  \MgII\ absorbers (Steidel 1993;
Steidel et al. 1994) and the deep imaging survey of Le Brun et al. (1993).
Over the redshift interval $0.2<z<1$, Steidel et al. (1994) find no
signifiant evolution in the space density, rest-frame B or K luminosity
or B$-$K color for the quasar absorption line-selected galaxies. 
Comparison between the predicted and observed sizes of galactic halos 
leads to a normalization of a Schechter-type luminosity function consistent
with the I-band  normalization factor derived by Lilly et al. (1995) for the
CFRS over the redshift range 0.2-0.5. Their value, as well as that estimated 
 by  Steidel et al. (1994) for their K-band  \MgII\ absorber sample, is
about a factor two larger  than the B-band  normalization factor of Loveday 
et al. (1992) for a local field galaxy sample.
Further analysis of Steidel et al.'s sample by Lilly et al. (1995)
shows a marginally significant increase in the average luminosity of
quasar absorption line-selected galaxies bluer than present-day Sbc.

To study the evolution of the field galaxy population using samples selected
with the same criterion, we have extended our survey of quasar absortion
line-selected galaxies up to $\za\sim$ 1.3-1.5. We aim at determining the
evolution of the galaxy spectral properties, continuum emission and rest-frame
equivalent width of the \OII$\lambda 3727$  emission line, over a
large redshift interval to ascertain the evolution in star formation activity.
Other properties, such as their kinematics, physical conditions and chemical
aboundances in the halo phase, could be obtained from the analysis of
high spectral resolution spectra of the quasars. Results for two fields
(Bergeron et al. 1992) suggest that \MgII\ absorption-selected
galaxies at $0.7<z<1.1$ have properties similar to those at
$\left<z\right>\sim 0.4$.

The observations and data reduction are outlined in Sect. 2. The properties 
of the \MgII\ absorber samples are presented in Sect. 3. 
To study the  cosmological evolution in the star formation activity,
we have built templates of the energy distribution of the \MgII\
absorption-selected galaxies for two redshift intervals 0.15-0.7 and
0.7-1.3. From the analysis of these templates, presented  in Sect. 4, we
derive a rough estimate of the chemical abundances in the central regions of
the $z \sim 0.5$ absorbers and investigate the past history of star formation 
in the absorbers. In Sect. 5, we discuss the implications of our
results on the evolution of the luminosity, overall size and star formation 
activity of \MgII\ absorption-selected galaxies.


\section{Observations}

\subsection{Sample selection }

\begin{table}
\caption{The Quasars sample}
\label{TQSO}
\tablehead{
  \hline
  Quasar & $\alpha$ (B1950) & $\delta$ (B1950)          & \ze\\
         & (h m s)          & (\degr\ \arcmin\ \arcsec) &\\
  \hline
}
\tabletail{\hline}

\begin{flushleft}
\begin{MonTabular}{lccc}
0002$-$422 & 00 22 15 & $-$42 14 11 & 2.763\\
0102$-$190 & 01 02 49 & $-$19 02 46 & 3.035\\
0109+200   & 01 09 29 &   +20 04 27 & 0.746\\
0150$-$203 & 01 50 05 & $-$20 15 54 & 2.147\\
0151+045   & 01 51 52 &   +04 33 35 & 0.404\\
0207$-$003 & 02 07 17 & $-$00 19 15 & 2.849\\
0229+131   & 02 29 02 &   +13 09 41 & 2.065\\
0235+164   & 02 35 53 &   +16 24 04 & 0.904\\
0302$-$222 & 03 02 36 & $-$22 23 29 & 1.409\\
0334$-$204 & 03 34 13 & $-$20 29 30 & 3.130\\
0424$-$131 & 04 24 48 & $-$13 09 33 & 2.166\\
0454+039   & 04 54 09 &   +03 56 15 & 1.343\\
1038+064   & 10 38 41 &   +06 25 59 & 1.270\\
1101$-$264 & 11 01 00 & $-$26 29 05 & 2.145\\
1127$-$145 & 11 27 36 & $-$14 32 54 & 1.182\\
1209+107   & 12 09 08 &   +10 46 58 & 2.191\\
1332+552   & 13 32 16 &   +55 16 46 & 1.249\\
1511+103   & 15 11 04 &   +10 22 15 & 1.546\\
1556$-$245 & 15 56 41 & $-$24 34 11 & 2.815\\
2000$-$330 & 20 00 13 & $-$33 00 13 & 3.777\\
2128$-$123 & 21 28 53 & $-$12 20 21 & 0.501\\
2145+067   & 21 45 36 &   +06 43 41 & 0.990\\
2206$-$199 & 22 06 07 & $-$19 58 44 & 2.544\\
2248+192   & 22 48 06 &   +19 15 25 & 1.806\\
2357$-$348 & 23 57 06 & $-$34 52 04 & 2.070\\
\end{MonTabular}
\end{flushleft}
\end{table}

We have selected quasar fields with \MgII\ absorbers principally from
homogeneous absorption-line samples (Sargent et al. 1988a, 1988b and
1989; Barthel et al. 1990) available prior to the \MgII\ survey of
Steidel \& Sargent (1992). For a few quasars, other studies have been used
(Hunstead et al. 1986; Ulrich 1989; Petitjean \& Bergeron 1990).
The absorption redshifts are in the range $0.7<\za <1.3$ with a few 
exceptions up to $\za$=1.5. The rest-frame equivalent widths of the 
\MgII$\lambda$2796 absorption lines are larger than 0.6\AA\ for all the
absorbers but one (see Table 2). 

The technique for identifying the absorbers is the same as that adopted by
Bergeron \& Boiss\'e (1991): r broad-band imaging followed by
spectrophotometry of the absorber candidates. The absolute magnitude range
for $\za\sim$0.4 absorbers is $-23.0\le\Mr\le -19.5$ (Bergeron \& Boiss\'e
1991; Le Brun et al. 1993). At this redshift, the r band roughly coincides
with the redshifted B band. At higher redshifts, $\za\sim$0.8-1.0, this
corresponds to an apparent magnitude range of $25.3\le\mr\le 21.8$  and
\mr=23.7 for a $L^\star$ galaxy, ignoring the relative color-term correction
between these two redshift intervals. We have tried at first to get redshift
data for galaxies as faint as $\mr\sim$24-24.5. For such faint objects, the
spectra
obtained at ESO with the NTT and the EMMI spectrograph in the long-slit mode
with typical exposure times of 3 or 4.5 hours were too noisy to derive galaxy
redshifts. Consequently, we later concentrate on fields with absorber
candidates brighter than \mr=23.5 within 15\arcsec ~from the quasar image,
thus limiting the sample to the bright end of the absorber luminosity
function. Since compact galaxies at $\zg\sim 1$ may not be spatially
resolved at a typical seeing of FWHM$\simeq 1\arcsec$, spectroscopic
follow-up has been conducted for any object in the close neighbourhood of
the quasar image. Objects with $\mr\simeq 23.5$ can be detected down to
angular separations from $\mr\sim 17$ quasar sighlines of
$\theta\simeq 2\arcsec$, or $D$=22\h50 kpc at \zg=1.

\subsection{Observations and data Reduction }

The observing modes and data reduction steps were similar to those described
in Bergeron et al. (1992). Only deep imaging in the r band  
($\lambda_0$=6410\AA\ and  $\Delta\lambda$=1540\AA: close to r Cousins) 
is available for most fields and the observed magnitudes \mr\ given below 
are estimated in the standard Vega-based system.
Different grisms have been used to match the expected wavelength
range of the Balmer discontinuity and the Ca\,{\sc ii} break, including the
\OII$\lambda 3727$ emission line and the H$\beta$,
\forb O{iii}$\lambda 5007$ range when possible. The flux calibration of the
imaging and spectroscopic data have been obtained from the observations of
photometric and spectroscopic standard stars.

Data reduction and analysis were done with the ESO MIDAS reduction packages.
For the spectral calibration, we used the LONG SLIT context with the option
of a two dimension calibration with usually a third degree polynomial in each
direction. For each object, the sky is then fitted with a polynomial of degree
one or two for each column. The spectrum of each galaxy is a Gaussian weighted
sum of the rows of the object. The spectra were obtained using a slit width
of 1.5$\arcsec$ or 1.8$\arcsec$ and a 230\AA\ mm$^{-1}$ grism, resulting in a
spectral resolution of FWHM=15 or 16.5\AA.

To estimate the absolute rest-frame B magnitudes, we have derived
$k$-corrections with the help of the spectral energy distributions (SED) 
given for local galaxies by Coleman et al. (1980). For the redshift 
range of our sample, selection of the rest-frame B band (Johnson filter: 
$\lambda_0$=4420\AA, $\Delta\lambda$=950\AA) instead of the rest-frame r band 
minimizes the  magnitude and uncertainty of $k$-corrections. 
Our sample can thus also be more directly compared with local-galaxy samples 
and higher redshift ones as that of Steidel et al. (1994) and the CFRS 
(see e.g. Lilly et al. 1995). 
For each galaxy, we have assigned a spectral type by comparing its rest-frame
SED with those given by Coleman et al. (1980) using the
strength of the Ba+Ca\,{\sc ii} break, or of only the Ba break when our
observed range does not extend far enough in the red. The Ba+Ca\,{\sc ii}
break has been estimated using continuum values averaged over the
rest-wavelength intervals 4050-4250\AA\ and 3400-3600\AA, whereas for the
Ca\,{\sc ii} break only the selected intervals are 4050-4250\AA\ and
3750-3900\AA. We first express the observed magnitudes in the AB system, 
the shift from Vega-type to AB magnitudes being equal to 
$m_{\mathrm AB}({\mathrm r}) = m({\mathrm r}) + 0.19$.
The $k$-correction is then computed following the procedure outlined by Lilly et al. (1995:
 the first equation in their Sect. 2.1). The $k$-correction color term (B$-$r$_{\rm rest}$) 
is estimated in the AB system from a comparison between two spectral regions 
of the tabulated SED for a galaxy of same morphological type as the absorber: 
firstly, the region of the galaxy rest-frame spectrum corresponding to the observed 
r-band wavelength range and, secondly, the B-band  rest-frame wavelength region. 

\subsection{Profile subtraction}

To probe further the fields without a candidate absorber, we have used an
empirical Point Spread Function (PSF) for profile
subtraction to the quasar image. The PSF has been constructed by averaging the
bright stars in the whole field. To exclude the saturated stars, we have
taken into account only the objects with a FWHM equal to the
resolution in both directions. Depending on the field, the number of
stars included in the PSF ranges from $\simeq~$5 to 15.
A median average of all the star images has been
built after normalization of the peak intensity of each stellar image and
recentering. At radial distances larger than 3 FWHM, the background has then
been smoothed using a Gaussian filter with a $\sigma$ of 3 pixels.

Outside the quasar images, the 3$\sigma$ detection limit varies from
\mr = 24.3 to 25.2 and objects at this magnitude limit could have been
detected on the raw images down to radial distances from the quasar center of
2.5 to 3.0$\arcsec$. After PSF subtraction to the quasar image, this
minimum radial distance decreases to 1.1 to 1.5 $\arcsec$ for similar or
slightly brighter ($\Delta\mr=-$0.3) magnitudes.

\begin{table*}
\caption{The $0.7<\za<1.3$ quasar \MgII\ absorption-selected galaxies}
\label{Thigh}
\tablehead{
  \hline
  Field & \za & \Wr(\MgII)$^a$ & \zg & $\Delta\alpha$ & $\Delta\delta$ &
  $\theta$ & $D$ & \mr & \Mab(B) & \Wr(\OII) & Break                       &
  galaxy\\
        &     & \AA            &     & \arcsec        & \arcsec        &
  \arcsec  & kpc &     &         & \AA       & {\scriptsize Ba+Ca{\sc ii}} &
  type\\
  \hline
}
\tabletail{\hline}
\begin{flushleft}
\begin{MonTabular}{lcccrrccclrcc}
0002$-$422 & 0.8363 & 4.68,4.03 & 0.840 &  $-$6.4 & $-$3.4 &  7.1 &  73.1 & 22.6 & $-$22.47 &
           $\leq$12.0 & 2.8 & E\\
0102$-$190 & 1.0262 & 0.67,0.69 & 1.025 &  $-$0.7 & $-$4.9 &  5.0 &  55.0 & 22.9 & $-$22.21 &
           54.0 & 1.5 & Scd\\
0302$-$222 & 1.0095 & 1.16,0.96 & 1.000 &  $-$2.6 & $-$7.2 &  7.7 &  84.0 & 23.1 & $-$22.80 &
            $\leq$10.0 & 3.3 & E\\
0334$-$204 & 1.1174 & 2.06,1.75 & 1.120 &     2.9 & $-$7.3 &  7.9 &  89.0 & 22.6 & $-$23.30 &
           91.0 & 1.7 & Sbc\\
1556$-$245 & 0.7713 & 2.07,1.91 & 0.769 &  $-$3.0 &    4.7 &  5.6 &  55.4 & 22.7 & $-$21.89 &
           25.0 & 2.7 & E\\
           &        &           & 0.771 &  $-$6.0 &    4.5 &  7.5 &  74.3 & 21.4 & $-$22.56 &
           30.0 & 1.6 & Scd\\
2000$-$330 & 0.7917 & 1.56,1.23 & 0.791 &     1.2 & $-$6.6 &  6.7 &  67.1 & 21.6 & $-$22.47 &
           53.0 & 1.5 & Scd\\
2145+067   & 0.7908 & 0.61,0.46 & 0.790 &     0.4 & $-$5.5 &  5.5 &  55.0 & 22.5 & $-$21.29 &
           16.7 & 1.03$^b$ & Im\\
2206$-$199 & 0.7520 & 0.93,0.77 & 0.755 &     2.4 &    5.4 &  6.0 &  59.0 & 22.6 & $-$20.98 &
           10.0 & 1.0 & Im\\
           & 0.9482 & 0.41,0.21 & 0.948 &  $-$6.4 &    8.6 & 10.9 & 117.  & 21.9 & $-$22.56 &
            5.0 & 1.04 & Im\\
           & 1.0169 & 0.93,0.95 & 1.017 & $-$12.6 & $-$4.9 & 13.5 & 148.  & 21.0 & $-$23.69 &
           64.0 & 1.18 & Im\\
\end{MonTabular}
\footnotesize
\\\smallskip
$^a$~$\lambda\lambda$2796,2803 \\
$^b$~Ba Break only (continuum values averaged over the
rest-wavelength intervals 3750-3900\AA\ and 3400-3600\AA)\\
\end{flushleft}
\end{table*}


\section{The galaxy samples }

Imaging and spectroscopic results are presented for 13 fields comprising
16 \MgII\ absorbers at $0.7<\za<1.3$, excluding one damped \Lya\
system at \za=0.8597 toward Q 0454+039 (Steidel et al. 1995; Le Brun
et al. 1997), and three lower redshift, $\za<0.7$, \MgII\ absorbers. We
have spectroscopic information on usually the two objects closest from the
quasar image when one \MgII\ absorber is known, and two or three
galaxies further away. The faintest galaxy for which a redshift has been
successfully determined is in the 0334$-$204 field, at \zg=1.220 and with
\mr=23.8 (see Table \ref{Tfirst}).

\subsection{The high-redshift absorber sample }

\begin{table*}
\caption{
  High redshift candidate absorbers within 15$\arcsec$ from the quasar
  sightline
}
\label{Tfirst}
\tablehead{
  \hline
  Field & \za & \Wr(\MgII) & \zg & $\Delta\alpha$ & $\Delta\delta$ &
  $\theta$ & $D$ & \mr & \Mab(B) & \Wr(\OII) & Break &galaxy\\
        &     & \AA        &     & \arcsec        & \arcsec        &
  \arcsec  & kpc &     &         & \AA       & Ba + Ca{\sc ii}& type\\
  \hline
}
\tabletail{\hline}

\begin{flushleft}
\begin{MonTabular}{lcccrrccclrcc}
0150$-$203 & 0.7800 & 0.36,0.21 & 0.603 &   5.8  & $-$5.7 &  8.2 & 72.9 & 20.7 & $-$22.31 &
           17. & 1.4 & Scd\\
0207$-$003 & 1.0447 & 0.67,0.53 & 0.366 & $-$0.9 & $-$4.3 &  4.4 & 29.7 & 22.0 & $-$19.43 &
           25. & 0.90 &Im\\
           &        &           & 0.703 &    7.8 & $-$6.8 & 10.4 & 99.2 & 23.5 & $-$20.09 &
           33. & 1.5 & Scd\\
0302$-$222 & 1.0095 & 1.16,0.96 & 0.663 & $-$8.5 &   11.9 & 14.2 & 132. & 21.2 & $-$22.16 &
            9. & 1.1$^d$ &Scd\\
0334$-$204 & 1.4890 &           & -     & $-$8.8 & $-$2.8 &  9.3 & 113.$^a$ & 22.9 & $-$23.55$^a$ &
            -  & - &Scd \\
           &        &           & 1.22  &    1.9 & $-$12. & 12.6 & 146. & 23.8 & $-$23.36 &
           31. & 2.5 & E\\
0424$-$131 & 1.0345 & 0.99,0.90 & -     &    6.3 &    4.6 &  7.8 & 86.1$^a$ & 21.5 & $-$23.64$^a$ &
           $\le$6.$^b$ & 1.2$^c$ & Scd\\
           &        &           & -     & $-$8.8 &    0.7 &  8.7 & 96.0$^a$ & 22.5 & $-$22.24$^a$ &
           $\le$37.$^b$ & 1.0$^c$& Im\\
2000$-$330 & 1.4542 & 0.19,0.11 & -     &   7.0 &  $-$2.6 &  7.4 & 89.8$^a$ & 22.5 & $-$23.20$^a$ &
           $\le$12.$^b$ & 1.0$^c$& Im\\
2248+192   & 1.2701 & 1.03,0.68 & 0.771 &    5.4 & $-$8.9 & 10.6 & 105. & 22.6 & $-$21.10 &
           42. & 1.2& Im\\
2357$-$348 & 0.995  &           & -     &    2.1 & $-$2.7 &  3.6 & 39.2$^a$ & 23.4 & $-$21.22$^a$ &
           $\le$20.$^b$ & 1.0 & Im\\
           &        &           & 0.415 &    7.3 & $-$4.9 &  8.8 & 64.0 & 21.3 & $-$20.46 &
            9.$^e$  & - & Scd\\
\end{MonTabular}
\\\smallskip
$^a$~we have assumed \zg=\za\ for inconclusive spectra\\
$^b$~3$\sigma$\ limit\\
$^c$~Ba Break only, as defined in Table \ref{Thigh}\\
$^d$~Ca\,{\sc ii} Break only (continuum values averaged over the
    rest-wavelength intervals 4050-4250\AA\ and 3750-3900\AA)\\
$^e$ \Wr(\OIII$\lambda$5007)\\
\end{flushleft}
\end{table*}

We have identified 11 of the 16 \MgII\ absorption-selected galaxies at
$0.7<\zg<1.3$, including already known absorbers (Bergeron \& Boiss\'e 1991;
Bergeron et al. 1992). These galaxies are associated with only 10 \MgII\
absorption systems, as in the 1556$-$245 field two galaxies with similar impact
parameters are at the same redshift and can both contribute to the observed
\MgII\ absorption. The results are presented in Table \ref{Thigh} and
the inconclusive cases are discussed in Sect. 3.2.
The 2206$-$199 field was reobserved to identify the third aborber at
\za=0.9482 and obtain spectrophotometric information not previously available.
The 2000$-$330 field was observed with the aim to identify the \za=1.4542
\MgII\ absorber. The galaxy closest to the quasar image is at
\zg=0.791, and we have identified a posteriori in the \Lya\ forest region of
this quasar (Table 3 in Hunstead et al. 1986) a \MgII\ doublet at
\za=0.7918 with \Wr(\MgII$\lambda\lambda 2796,2803$)=1.6 and 1.2 \AA.
We favor this identification over that given by Hunstead et al.
who identified these lines with \Lya. Although they also mentioned
possible associated \Lyb\ absorption, we note that the latter are much
stronger than their tentative \Lya\ counterpart and the redshift
agreement between the two sets of lines, if from the Lyman series, is poor
(of only 0.002-0.003). The \za=1.0095 multiple \MgII\ system in Q 0302$-$222
has three components spanning 170 \kms (Petitjean \& Bergeron 1990). Two
components have an associated strong \FeII\ absorption, typical of absorbers
with large \HI\ column densities, N(\HI) $>$ a few 10$^{19}$ cm$^{-2}$
(Bergeron \& Stasi\'nska 1986). This system may arise from two galaxies,
the one identified in this work and a second one
much closer to the quasar sightline (Lebrun et al. 1997).
  
\begin{figure}
  \psfig{figure=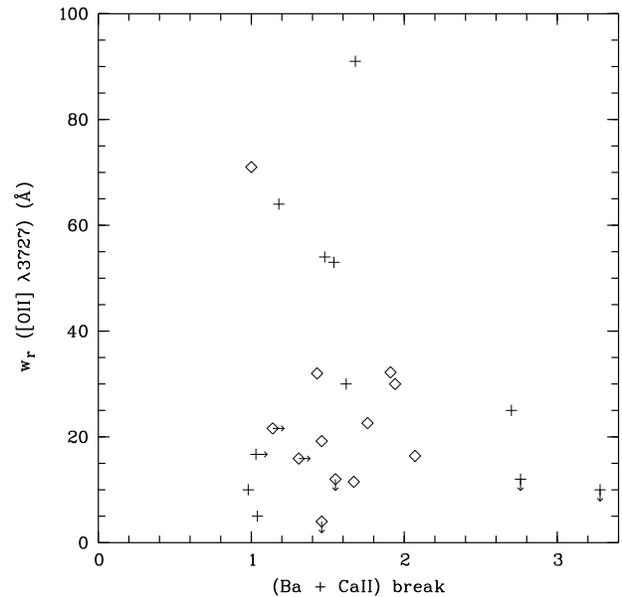,height=8cm}
  \caption[ ]{
    The rest-frame equivalent width of \OII$\lambda 3727$
    versus (Ba + Ca\,{\sc ii}) break. The absorbers at $\zg >$ 0.7 and
    $<$ 0.7 are shown as crosses and diamonds respectively. Cases with data
    available for only either the Ba or the Ca\,{\sc ii} break are
    represented by  horizontal arrows and the vertical arrows
    indicate \Wr\ upper limits
}
  \label{FWDisc}
\end{figure}

The high-redshift galaxy sample covers the observed magnitude range
$21.0\le\mr\le 23.1$, leading to $k$-corrected B absolute magnitudes of
$-23.7\le\Mab$(B)$\le -21.0$ at $0.755\le\zg\le 1.120$. All the
galaxies have luminosities similar or brighter than present day $L^{\star}$
galaxies (see Table \ref{Thigh}). Their spectral types cover the whole range
from elliptical to irregular galaxies with however a predominence of
late type galaxies (Scd and Im). The
spectral types have been assigned using the Ba + Ca\,{\sc ii} break, the
Ca\,{\sc ii} break tracing the cooler stellar population. Another alternative
would have been to use the rest-frame equivalent width of the
\OII$\lambda 3727$ emission line. The latter is however more representative
of the young stellar population, whose contribution to the galaxy energy
distribution is predominant in the UV range. The lack of an anti-correlation
between the strength of the Ba + Ca\,{\sc ii} break and \Wr(\OII$\lambda 3727$)
(see Fig. \ref{FWDisc}), as confirmed by the Spearman or Kendall rank
correlation tests, indicates the presence of a young stellar population
in excess of that observed in present-day galaxies of the same spectral type.
This phenomenon is confirmed by the spectral shape of the energy distribution
bluewards of the Balmer discontinuity of the absorber template spectrum
presented in Sect. 4. We note however that all the
objects with strongest \OII\ emission are in  Fig. \ref{FWDisc} among the
blue half of the sample.

\subsection{Closest neighbours and inconclusives cases }

Among the remaining seven fields with non-identified, high-redshift \MgII\
absorbers, information on  galaxies closest ($\theta<15\arcsec$) to the 
quasar image are given in Table \ref{Tfirst}, as well as another galaxy in
the 0302$-$222 field. This large upper limit of $\theta$ has been selected 
so as to ensure that even the brightest possible absorber ($\Mab$(B)$\simeq
-24$) are included in the search sample.
Five galaxies have been identified, of which
three at $\zg >$ 0.7, together with two lower redshift absorbers. The latter
are presented in Sect. 3.3 and included in Table \ref{Tlow}.

For four of the seven non-identified cases, the galaxy spectra are
inconclusive. These intrinsically bright galaxies could be the \MgII\
absorber, since their luminosity and halo radius would be consistent with the
scaling law found for the identified \MgII\ absorbers.

\begin{figure}
  \psfig{figure=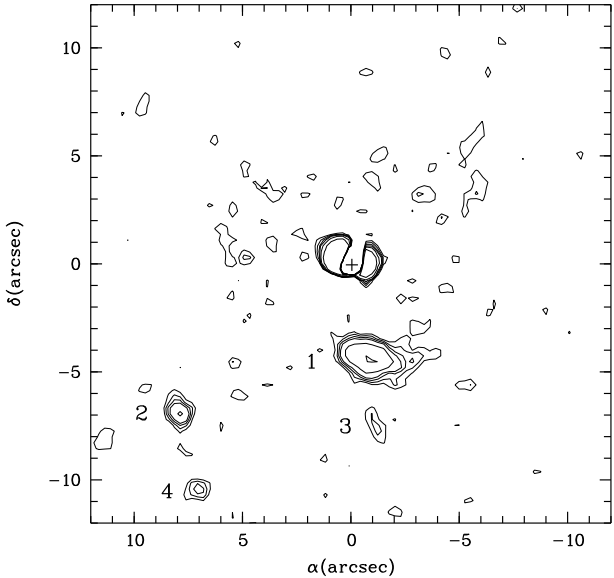,height=8cm}
  \caption[ ]{
    Close vicinity of Q 0207$-$003 after PSF subtraction and smoothing.
    Contour level values are 2,3,4,5,8 and 16$\sigma$
    }
  \label{sub1}
\end{figure}

\begin{figure}
  \psfig{figure=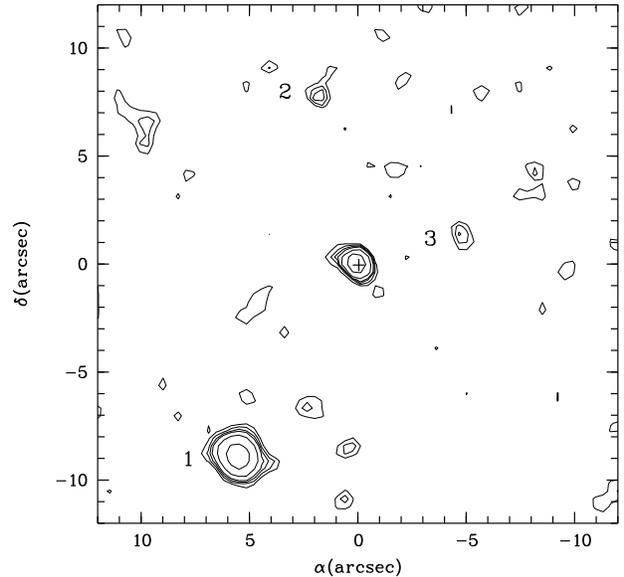,height=8cm}
  \caption[ ]{
    Close vicinity of Q 2248+192 after PSF subtraction and smoothing.
    Contour level values are 2,3,4,5,8 and 16$\sigma$
    }
  \label{sub2}
\end{figure}

For the other three cases, we have searched for objects fainter than the
identified \zg $<$ \za\ galaxies within 10$\arcsec$ from the quasar image.
In the 0150$-$203 field, there is a galaxy at an impact parameter of
5.3$\arcsec$ with \mr=22.0 for which we do not have spectroscopic data.
In 0207$-$003 field, there is one object at  $\theta =7.4\arcsec$
(object 3 in Fig. \ref{sub1}) with a magnitude fainter than
\mr(3$\sigma$)=24.3 which is also detected in the deeper (\mr(3$\sigma$)=25.2)
imaging survey of Le Brun et al. (1993) with \mr=24.9 (object 4 in Fig.
\ref{sub1} is also detected by Le Brun et al.).
In the 2248+192 field, there is one object at $\theta = 8.0\arcsec$ with a
magnitude \mr$\simeq$\mr(3$\sigma$)=25.0 (object 2 in Fig. \ref{sub2}) and a
fainter one at $\theta = 5.0\arcsec$ (object 3 in Fig. \ref{sub2}) which
could be a noise artefact.
There is no additional object brigther than our 3$\sigma$ limit
after PSF subtraction to the quasar image. Results for the  0207$-$003 and
2248+192 fields are presented in Figs. \ref{sub1} and \ref{sub2} after a
Gaussian smoothing with a width equal to half the spatial resolution.
Objects 1 and 2 in Fig. \ref{sub1} and object 1 in Fig. \ref{sub2} are those
listed in Table \ref{Tfirst}.
In the three fields, the faint objects without spectroscopic information
would have magnitude well within the range found for the identified
absorbers, if they were at \zg=\za.

\begin{table*}
\caption{The $0.15<\za<0.7$ quasar \MgII\ absorption-selected galaxies}
\label{Tlow}
\tablehead{
  \hline
  Field & \za & \Wr(\MgII) & \zg & $\Delta\alpha$ & $\Delta\delta$ &
  $\theta$ & $D$ & \mr & \Mab(B) & \Wr(\OII) & Break                       &
  galaxy\\
        &     & \AA            &     & \arcsec        & \arcsec        &
  \arcsec  & kpc &     &         & \AA       & {\scriptsize Ba+Ca{\sc ii}} &
  type\\
  \hline
}
\tabletail{\hline}

\begin{flushleft}
\begin{MonTabular}{lcccrrccclrcc}
0109+200   & 0.5346 & 2.26,1.71 & 0.534 &    0.5 &     7.0 &  7.1 &  59.4 &
           21.7 & $-$20.86 &  $\leq$4.0 & 1.5 & Scd\\
0150$-$203 & 0.3887 & 0.58,0.32 & 0.383 &    7.7 &   7.2   & 10.5 &  72.9 &
           21.3 & $-$20.23 & 30.0 & 1.9 & Sbc\\
0151+045   & 0.1602 & 1.55,1.55 & 0.160 & $-$6.2 &  $-$1.7 &  6.4 &  23.9 &
           19.1 & $-$20.17 & 15.9 & 1.3$^c$ & Sbc\\
           &        &           & 0.160 & $-$3.0 &    10.5 & 10.9 &  40.7 &
           20.2 & $-$19.14 & 21.6 & 1.1$^c$ & Scd\\
0229+131   & 0.4176 & 0.67,0.75 & 0.417 & $-$5.6 &     3.9 &  6.8 &  49.7 &
           20.5 & $-$21.27 & 22.6 & 1.8 & Sbc\\
0235+164   & 0.524  & 2.42,2.34 & 0.524 &    0.3 &  $-$1.9 &  1.9 &  15.7 &
           21.2 & $-$21.21 & 25.0 &      & Im\\
0302$-$222$^a$ & 0.419 & 0.90,0.90      & 0.418 & $-$18. &  $-$15. & 23.0 & 168.  &
           18.4 & $-$23.38 & 32.0 & 1.4 & Scd\\
0454+039   & 0.072  & 0.72,0.65 & 0.072 &    3.8 &     1.4 &  4.0 &   7.6 &
           20.5 & $-$17.02 & 16.0$^b$& 1.5 & Scd\\
1038+064   & 0.442  & 0.30,0.25 & 0.441 &    9.2 &     2.1 &  9.4 &  70.9 &
           21.2 & $-$20.75 & 16.4 & 2.1 & Sbc\\
1101$-$264 & 0.3591 & 0.49,0.40 & 0.359 &    5.7 & $-$10.8 & 12.2 &  81.4 &
            20.4 & $-$20.95 & 8.0 &\     & Sbc\\
1127$-$145 & 0.3129 & 2.21,1.90 & 0.313 &    8.7 &     3.9 &  9.6 &  58.7 &
           19.5 & $-$21.47 & 32.2 & 1.9 & Sbc\\
1209+107   & 0.3930 & 1.00,0.54 & 0.392 &    5.1 &     4.9 &  7.1 &  50.0 &
           21.9 & $-$19.70 & 71.0 & 1.0 & Im\\
1332+552   & 0.374  & 2.90,2.90 & 0.373 & $-$2.5 &     4.4 &  5.0 &  34.2 &
           20.7 & $-$20.75 &  $\leq$12.0 & 1.5 & Sbc\\
1511+103   & 0.4369 & 0.45,0.35 & 0.437 &    5.1 &     4.6 &  6.9 &  51.8 &
           21.6 & $-$20.32 & 19.2 & 1.5 & Scd\\
2128$-$123 & 0.4299 & 0.40,0.37 & 0.430 &    6.7 &     5.4 &  8.6 &  63.9 &
           21.0 & $-$20.87 & 11.5 & 1.7 & Sbc\\
\end{MonTabular}
\footnotesize
\\\smallskip
$^a$~ although very far away, this galaxy is very bright and follows the D,\Mr\ correlation\\
$^b$ \Wr(H$\alpha$)\\
$^c$ Ca\,{\sc ii} Break only, as defined in Table \ref{Thigh}\\
\end{flushleft}
\end{table*}

\begin{figure}
  \psfig{figure=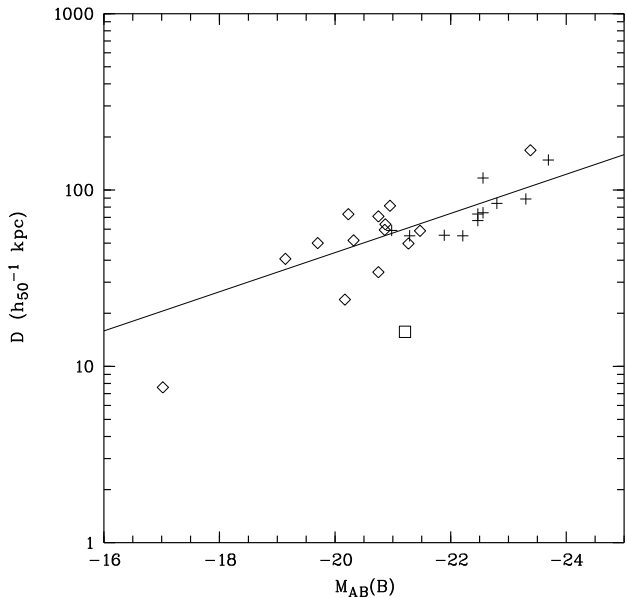,height=8cm}
  \caption[ ]{
    The scaling-law absorber radius versus rest-frame \Mab(B) magnitude.
    The symbols are the same as in the Fig. \ref{FWDisc} and the squares
    represent damped \Lya/21 cm absorbers. The line shown is the best fit
    of the data
    }
  \label{FDMb}
\end{figure}

We now investigate whether there are interlopers among the five close
neighbour galaxies listed in Table \ref{Tfirst}. The three at $\zg >$ 0.7
have impact parameters larger than found for \MgII\ absorbers (see the
scaling-law radius-luminosity presented for the absorbers in
Fig. \ref{FDMb}), whereas the two at $\zg <$ 0.7
have impacts parameters within the range expected for \MgII\ absorbers. For
the $\zg$=0.366 galaxy toward Q 0207$-$003, there are no published data of the
quasar spectrum in the expected wavelength range of the associated
\MgII\ absorption doublet.  For the $\zg$=0.603 galaxy toward
Q 0150$-$203, the  expected associated \MgII\ absorption doublet is not
detected at the 4$\sigma$ level with \Wr(\MgII$\lambda$2796 or
$\lambda$2803) $<$ 0.16\AA\ (Sargent et al. 1988a). The lack of \MgII\
absorption does not imply an absence of extended gaseous halo. Indeed, the
{\it HST} Key Project on quasar UV absorption line survey has revealed the
existence of low-redshift absorption systems of high ionization (see e.g.
Bergeron et al. 1994).

\subsection{The low-redshift absorber sample}

The sample of 15 \MgII\ absorption-selected galaxies listed in Table
\ref{Tlow} comprises (i) the 12 objects at $z<0.7$ studied by
Bergeron \& Boiss\'e  (1991), (ii) three new identified absorbers. Among the
latter, there is the low-redshift ($z=$0.072) dwarf galaxy discovered by
Steidel et al. (1993) in the field of Q 0454+039, for which we also had
obtained a red spectrum. There is only one strong emission line in our
observed spectral range; the energy distribution is very flat over the
interval 5800-7800\AA\ and the flux decreases redward of 7800\AA. Detailed
spectroscopic information on the \OII$\lambda 3727$
emission line was not given by Steidel et al., and the
rest-equivalent width listed in Table \ref{Tlow} refers to H$\alpha$. There
is a \MgII\ absorption doublet in the {\it HST-FOS} spectrum of the quasar
associated with this dwarf galaxy (Boiss\'e et al. 1997). The $z$=0.383 \MgII\
absorber toward Q 0150$-$203 has a magnitude close to that of a $L^{\star}$
galaxy and its impact parameter follows the scaling-law
$R=R^{\star}(l/L^{\star})^{0.3}$ found for \MgII\ absorbers
(Bergeron \& Boiss\'e 1991; Le Brun et al. 1993; Steidel 1993). The physical
properties of the $z$=0.418 \MgII\ absorber toward Q 0302$-$222 are
more extreme: it is about as bright as the brightest $z >$0.7 \MgII\
absorbers (\zg=1.017 toward Q 2206$-$199) and with an impact parameter
slightly larger. There are two other galaxies brighter than  \mr=22 closer to
the quasar sightline detected in {\it HST-PC} images (Le Brun et al. 1997);
 both are fairly red, $\mb-\mr>$2.5, and of elliptical morphology, thus
not as likely to be responsible for the \za=0.418 \MgII\ absorption.

The low-redshift galaxy sample covers the observed magnitude range
$18.4\le \mr\le 21.9$, leading to $k$-corrected B absolute magnitudes of
$-23.4\le\Mab$(B)$\le -17.0$ (or $-$19.1 when excluding the
low-redshift dwarf galaxy) at $0.072\le\zg\le 0.534$. This magnitude range
is roughly centered on the luminosity of a $L^{\star}$ galaxy, and fully
overlap with the smaller range found at higher-redshift (limited to galaxies
brighter than about  $L^{\star}$).
The spectral type have been assigned using either the Ba + Ca\,{\sc ii} break
or only the Ca\,{\sc ii} break. As for the high-redshift galaxy sample, they
cover a wide range of spectroscopic types, but with a
predominence of intermediate (Sbc and Scd) instead of late type galaxies .

\section{Stellar populations and metal abundances}

\subsection{Absorber templates}

\begin{figure}
  \psfig{figure=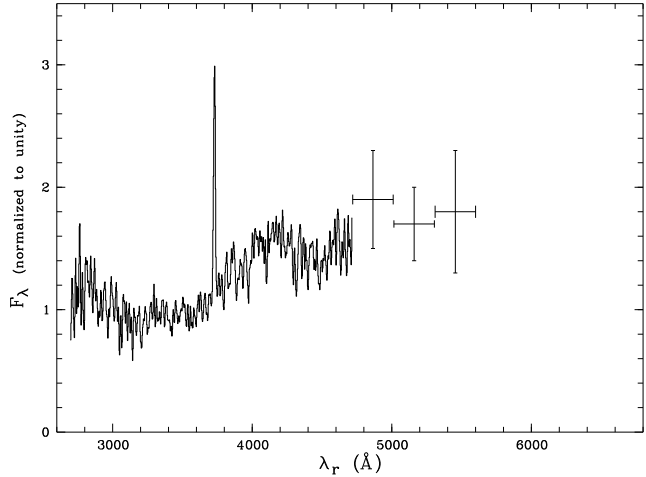,height=7cm}
  \caption[ ]{
    The template spectrum of the high-redshift absorbers. The crosses at the
    red end of the spectrum give the median value of the spectrum at these
    wavelength bins, the vertical bars representing 1$\sigma$ rms
    }
  \label{Fhightempl}
\end{figure}

We have built a template spectrum from the 11 high-redshift \MgII\
absorbers. Each rest-frame spectrum has been normalized to unity at the
continuum level of the observed or expected \OII$\lambda 3727$
emission line. In each overlaping wavelength
range, the available spectra have been averaged. The resulting template is
shown in Fig. \ref{Fhightempl}. A soft smoothing has been applied using
a Gaussian filter of FWHM=6.5\AA\ or 0.5 times the spectral resolution.
At the red end, the signal-to-noise ratio is very low and continuum values
averaged over $\sim 250$\AA\  are given together with $1\sigma$ rms error
bars. The rest-frame equivalent width of \OII$\lambda 3727$ equals
\Wr=30.0$\pm$ 1.3\AA\ (1$\sigma$ rms: an additional uncertainty of about
$\pm$2\AA\ arises from positioning of the continuum level) at
$\left<\zg\right>$=0.89. The spread in the individual values of
\Wr(\OII$\lambda 3727$) is very large from 5 to 91\AA.
The Ba and Ca\,{\sc ii} breaks are equal to 1.35 and 1.23 respectively. 
The SED is intermediate between those of
present-day Sbc and Scd galaxies redwards of the Balmer discontinuity, but
much bluer at shorter wavelengths. The flux of local Sbc and Scd galaxies is
steadily decreasing from 3500 to 2800\AA\ and the flux ratio
$F_{\lambda}(\lambda_{\mathrm{r}}2800)/F_{\lambda}(\lambda_{\mathrm{r}}3500)$
equals 0.62 and 0.89 for Sbc and Scd galaxies respectively (Coleman et
al. 1980). The corresponding value for the high-redshift template is 1.25,
similar to that of Im galaxies. This UV excess reveals the presence of
a much stronger star formation activity than in present-day galaxies.

\begin{figure}
  \psfig{figure=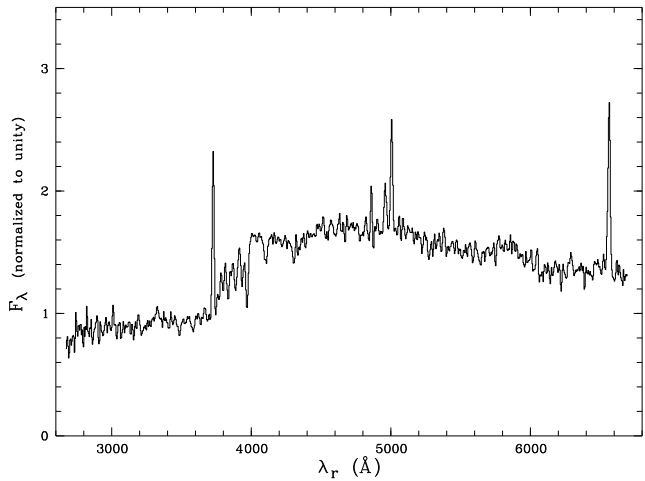,height=7cm}
  \caption[ ]{The template spectrum of the low-redshift absorbers}
  \label{Flowtempl}
\end{figure}

\begin{table}
\caption{Spectral properties of the absorber templates}
\label{Ttempl}
\tablehead{
  \hline
  Line &\multicolumn{4}{c}{Template}\\
  \cline{2-5}
       & \multicolumn{2}{c}{low \zg} & \multicolumn{2}{c}{high \zg}\\
       & blue & red & blue & red\\
  \hline
}
\tabletail{\hline}

\begin{flushleft}
\begin{MonTabular}{lcccc}
$\left<\zg\right>$           &  0.28 & 0.39 & 0.91 & 0.90\\
Ba + Ca\,{\sc ii} Break      & 1.48 & 1.90 & 1.21 & 2.10\\
$F_{\lambda}(\lambda_{\mathrm{r}}2800)/F_{\lambda}(\lambda_{\mathrm{r}}3500)$
 & 1.11 & 0.90 & 1.57 & 1.15\\
\Wr(\OII$\lambda 3727$)(\AA)   & 21.5 & 19.0 & 32.0 & 29.5\\
I(H$\beta$)$^a$                  &  0.35 & 0.34&&\\ 
I(\forb O{iii}$\lambda$5007)$^a$ &  0.49 &  0.90&&\\
I(H$\alpha$)$^a$                 &  1.10 & -&&\\     
I(\forb N{ii}$\lambda$6583)$^a$  & $\leq0.25^b$ & -&&\\
\end{MonTabular}
\footnotesize
\\\smallskip
$^a$ the line intensities are normalized to I(\OII$\lambda$3727)\\
$^b$ 4.5$\sigma$ limit\\
\end{flushleft}
\end{table}

Using the same procedure, we have built the template spectrum from the 15 
low-redshift \MgII\ absorbers and the resulting spectrum is shown in 
Fig. \ref{Flowtempl}. The rest-frame equivalent width of \OII$\lambda 3727$ 
equals \Wr=21.0$\pm$ 0.8\AA\ (the additional uncertainty  
 arising from positioning of the continuum level is $\pm$1\AA) at
$\left<\zg\right>$=0.36. This value is smaller than found for the
higher-redshift template by 30\%. The spread in the individual values of
\Wr(\OII$\lambda 3727$) is very large, from $<$4 to 71\AA\
(see Table \ref{Tlow}). The Ba and Ca\,{\sc ii} breaks are equal to 1.38 and
1.26 respectively, very close to the values found at higher redshift. The SED
is similar to that of a present-day Sbc galaxy redwards of the Balmer
discontinuity, and somewhat bluer at shorter wavelengths. The flux ratio
$F_{\lambda}(\lambda_{\mathrm{r}}2800)/F_{\lambda}(\lambda_{\mathrm{r}}3500)$
equals 0.90, very close to the value found for present-day Scd galaxies
(Coleman et al. 1980). This UV excess is weaker than found at higher redshift
but still reveals the presence of a star formation activity stronger than in
present-day galaxies.

\subsection{Stellar populations}

Although the templates are averaged spectra of individual galaxies of
different ages, we have compared them to stellar population synthesis models
to test the evolution in the star formation activity of the \MgII\
absorption-selected galaxies. We have used an updated version of the
spectral evolution models of stellar populations developed by Bruzual \&
Charlot (1993, 1996). These models combine the recent version of a
photometric model of isochrone synthesis with an updated library of stellar
spectra. The stellar populations have a solar metallicity and the star
formation rate is assumed to decrease exponentially, e$^{-t/tau}$, where
$tau$ equals 3, 5 and 9 Gyr for galaxies of morphological type Sa, Sb and Sc
respectively.

These models are used to investigate the past history of star formation in
the absorbing galaxies. The youngest stellar component is mainly traced by
the SED bluewards of the Balmer discontinuity, and the latter provides the
strongest constraint to the models. The low-redshift absorber template is
satisfactorily fitted with ages between 4.0,4.75 and 6.0 Gyr for Sa, Sb and
Sc morphological types respectively, but the UV excess at
$\lambda_{\mathrm{r}}<$3500\AA\ of the high-redshift one is not correctly
reproduced whatever the age of the stellar population or the galaxy
morphological type. This strongly suggests that the absorber templates 
comprise galaxies of too different ages.

\begin{figure}
  \psfig{figure=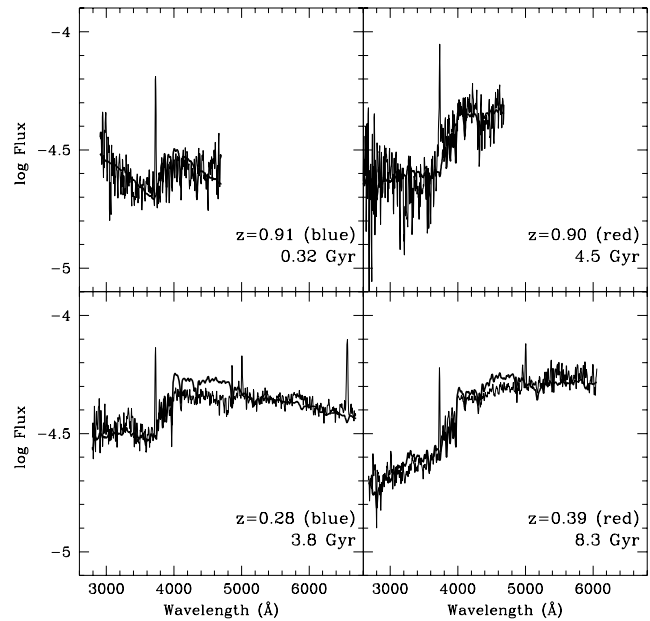,height=9cm}
  \caption[ ]{
    Modelisation of the stellar populations with different epochs of  
    burst of star formation for the red and blue populations of the 
    low- and high-redshift absorber templates 
    }
  \label{FSCt}
\end{figure}

\begin{figure}
  \psfig{figure=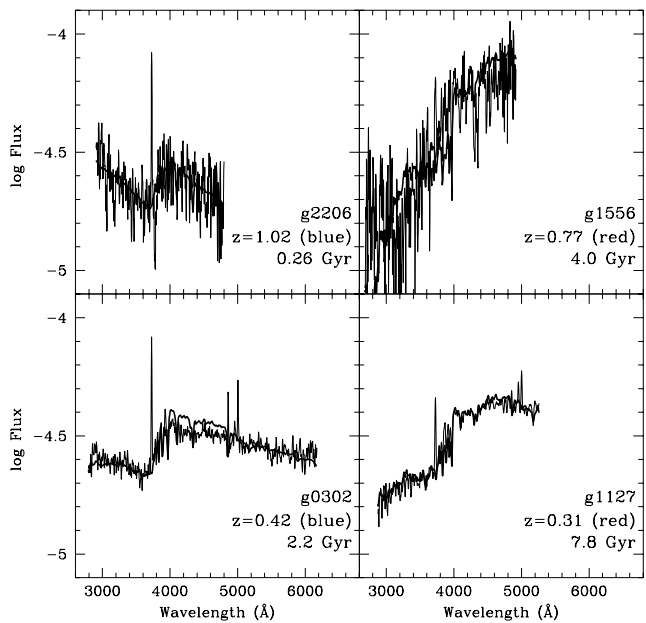,height=9cm}
  \caption[ ]{
    Modelisation of the stellar populations with different epochs of  
    burst of star formation for individual red and blue galaxies at
    low and high redshifts
    }
  \label{FSCg}
\end{figure}
 
At the suggestion of S. Charlot, we have then used more homogeneous
absorber subsamples differentiated by the value of their Ba + Ca\,{\sc ii}
break $\leogr 1.55$ (see Table \ref{Ttempl}). The results
of this analysis are shown in Fig. \ref{FSCt}. The red subtemplates
are well fitted by a Sb galaxy 4.5 and 8.3 Gyr old for $z\simeq$0.9 and 0.4
respectively. We note that the age difference of the stellar populations at
these two redshifts is about equal to the time elapsed between the two
redshifts for $h_{50}$=1 and  $q_0$ = 0, thus a similar galaxy formation epoch
for these two subclasses of absorbers. The best fit obtained for the blue
subtemplates gives for a Sc galaxy ages of 0.32 and 3.8 Gyr for $z\simeq$0.9 
and 0.3 respectively. For the higher redshift subtemplate spectra, the fits 
are fairly satisfactory whereas, for the lower redshift ones, there is some 
discrepency in the predicted SED in the wavelength range 4000-4800\AA,
especially for the blue subsample. 

To confirm these age estimates and check the reality of the discrepency in the 
4000-4800\AA\ region of the predicted SED, we have tried to fit individual galaxy 
spectra although their S/N is somewhat low especially at $z\simeq 0.9$. The resulting
fits are shown in  Fig. \ref{FSCg}. As for the low-redshift blue subtemplate, 
the predicted SED for G 0302$-$222 at $z$=0.418 does not correctly reproduce 
the observed spectrum around 4500\AA, which could be due to
some problem in the population synthesis models. The derived stellar population ages
(see  Fig. \ref{FSCg}) are very similar to those obtained with the higher S/N 
subtemplate spectra. This stengthens our conclusion that, at  $z\sim 1$, intense 
star formation activity is occuring in galaxies with a  Ba + Ca\,{\sc ii} break 
smaller than 1.55, i.e. about half of the Mg\,{\sc ii},  high-redshift absorber
subsample.

\subsection{Metal abundances}

To get a rough estimate of the  O/H and N/H abundances, we had to study the
template spectrum averaged over all low-redshift absorbers, whereas individual 
galaxy spectra could be considered for deriving solely  O/H. 

The \NII$\lambda$6568 emission line is not detected in the low-redshift
absorber template, although the  \OII$\lambda$3727 emission line is
stronger than  \OIII$\lambda$5007. This could be indicative of a relative
abundance N/O lower than the solar value. As our data are of too low
sensitivity and/or the gas temperature is too low to detect the weak
temperature-sensitive line \OIII$\lambda$4363, we have tried to get a
rough estimate of the mean abundances of the lower redshift absorbers using
the empirical methods developed by  Alloin et al. (1979) and  Pagel et al.
(1979). The empirical O/H abundance indicator proposed by the latter authors,
$R_{23}\equiv$(\OII$\lambda$3727+\OIII$\lambda\lambda$4959,5007)/H$\beta$,
has been widely adopted.

The O/H abundance cannot be derived unambiguously from the line ratio
$R_{23}$: the O/H,$R_{23}$ diagram has two branches with a turn-over region
at [O/H]$\simeq -$0.4. In the diagram of Edmunds \& Pagel (1984) and the
grid of \HII\ region models constructed by McGaugh (1991), the observed value
of $R_{23}$=5.0 falls within the turn-over region. The range of possible
values for [O/H] is $-1.3,-0.1$. Using the additional information provided
by the Oxygen line ratio
$O_{32}\equiv$\OIII$\lambda\lambda$4959,5007/\OII$\lambda$3727=0.9, and
the grid models of McGaugh (1991) leads to a the volume averaged
ionization parameter $\overline{U}\simeq$0.002, which narrows the range of
possible values values for [O/H] to $-1.0,-0.1$. The
\NII$\lambda$6568/\OII$\lambda$3727 line ratio can be used as a discriminant
between the upper and lower branches of the O/H,$R_{23}$ diagram, the
turn-over region ocurring around
log(\NII$\lambda$6568/\OII$\lambda$3727)$\approx -1$ (McGaugh 1994). For higher 
values of the latter, the \NII$\lambda$6568/\OII$\lambda$3727$,R_{23}$ diagram
is populated by normal spiral galaxies and for lower ones, low surface
brightness galaxies dominate. The \MgII\ absorption-selected galaxies are
normal spirals and the observed mean upper limit of
log(\NII$\lambda$6568/\OII$\lambda$3727) is equal to $-0.6$. This suggests that 
the above line intensity ratio should be larger that 0.4 times the measured
upper limit. The O/H abundance in the disk of \MgII\ absorption-selected
galaxies should then be in the range [O/H]=$-0.6,-0.1$.

We have also measured these line ratios for two individual galaxies:
G 0302$-$222 at \zg=0.418 and G 1127$-$145 at \zg=0.313. We then derived  
values of $R_{23}$=3.7 and 7.9 and $O_{32}$=0.7 and 1.0 respectively. These 
lead to  an ionization parameter $\overline{U}\simeq 0.001$ and 0.002. 
The possible ranges for [O/H] are $-1.1,-0.2$ and $-0.8,-0.4$. The latter is
well constrained as the observed value of $R_{23}$ is just at the turn-over
region. For both galaxies, the observed spectral range does not 
include the \NII$\lambda$6568 line. These results as well as those derived
from the low-redshift template are consistent with an underabundance factor 
of about three (within a factor two) at $\zg \sim$0.4.

\section{Discussion}

The overall sample of \MgII\ absorbing galaxies comprises 25(26) objects in
the redshift range (0.072)0.16$<z<$1.12 with absolute magnitudes
$-24 \leq\Mab(B)\leq -19.5(-17.2)$, the object of lowest redshift being
the only dwarf galaxy of our sample although absorbers of similar magnitude
could have been detected up to $z \sim 0.4$. There is only one clear case of
an interloper galaxy (G 0150$-$203 at $z$=0.603): its impact parameter is very close 
to the size of \MgII\
halos, as derived from the radius-luminosity scaling law, and UV spectroscopic
data are needed to investigate whether this galaxy has an extended
gaseous halo of high-ionization level.

The distribution of impact parameters $D$ does not show any evolution 
between the lower ($z<0.7$) and higher ($z>0.7$) redshift subsamples. 
The correlation between the halo radius $R$ and the galaxy luminosity 
previously discovered by
Bergeron \& Boiss\'e (1991) and Steidel (1993) still holds at
$\left<z\right> \simeq1$. The scaling law has been derived for the whole
sample, excluding however the 21cm absorber towards Q 0235+165, which should
arise in the galaxy disk, and the low-redshift dwarf galaxy which is a unique
case in our sample. The high- and low-redshift absorbers follow the same
correlation, although the luminosity of the former is on average higher as a
consequence of our limiting magnitude for spectroscopic identification.
The slope $\alpha$ of the scaling law,
$R/R^{\star}=(L_B/L_B^{\star})^{\alpha}$, is the same, $\alpha$=0.28, for
the overall fit and the upper envelope of the $D,L_B$ relationship. 
A somewhat shallower index, $\alpha$=0.20, has been derived by Steidel et al.
(1994) for their K-band selected sample.
Using \Mab$^{\star}$(B)$=-21.0$ (Loveday et al. 1992), we obtain for 
$D^{\star}$ and $R^{\star}$ values of  57 and 90 \h50\~kpc respectively. 
These two values differ by a factor of about 1.5 as would be expected if the 
geometry of the gaseous halos was spherical.

The evolution of the stellar component with cosmic time is given by the
variation of the  Ba + Ca\,{\sc ii} break for the older stellar population
and of the SED  bluewards of the Balmer discontinuity as well as 
\Wr(\OII$\lambda 3727$) for the young stellar population.
For the blue subsamples (Ba + Ca\,{\sc ii} break $<$ 1.55), the value of the  
Ba + Ca\,{\sc ii} break decreases by 22\% between
$\left<z\right>$ of 0.3 and 0.9, whereas \Wr(\OII$\lambda 3727$) increases
by 49\%. The latter is associated with an evolution of the SED
blueward of the Balmer discontinuity, with an increase of
$F_{\lambda}(\lambda_{\mathrm{r}}2800)/F_{\lambda}(\lambda_{\mathrm{r}}3500)$
by 41\%. For the red subsamples, the negative evolution in redshift of the  
Ba + Ca\,{\sc ii} break is probably not statistically significant 
due to the the very low S/N of the high-redshift subtemplate, especially just 
blueward of the Balmer discontinuity, whereas the blueing of the SED at 
 $\lambda_{\mathrm{r}} <$3500 \AA\ (increase of 
$F_{\lambda}(\lambda_{\mathrm{r}}2800)/F_{\lambda}(\lambda_{\mathrm{r}}3500)$ 
by 28\%) between $\left<z\right>$ of 0.4 and 0.9 is associated with an 
increase of \Wr(\OII$\lambda 3727$) by 55\%.

As shown in Sect. 4.3, these evolutions in redshift are consistent with passive
evolution for the red galaxy subsamples since the last epoch ($z\gg 1$) of strong
bursts of star formation, whereas for the blue subsamples they point towards an
intense stellar formation activity ocurring at $z\sim 1$.

\begin{figure}
  \psfig{figure=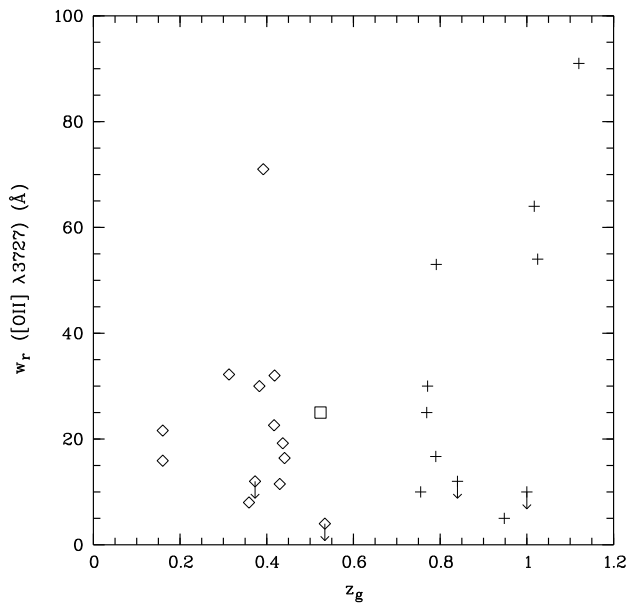,height=8cm}
  \caption[ ]{
    The rest-frame equivalent width of \OII$\lambda 3727$
    versus redshift and same symbols as in Fig. \ref{FDMb}
    }
  \label{FWZ}
\end{figure}

The evolution of \Wr(\OII$\lambda 3727$) with $z$ is not
unambiguously confirmed when analyzing the whole sample, which is due to
the large spread in the values of \Wr(\OII$\lambda 3727$) at both  low
and high redshifts (of about a factor of 18: see Fig.\ref{FWZ})
as well as the small  sample size.
 Applying the Spearman or Kendall rank correlation tests, we find that
the null hypothesis that \Wr(\OII$\lambda 3727$) and $z$ are uncorrelated can
be rejected at a level of only 47\%.
  
The ESO-Sculptor survey of a magnitude-selected sample, \mr $<$20.5, does not
reveal a clear evolution in redshift of \Wr(\OII$\lambda 3727$) for
$0.1 \le z \le 0.5$ (Galaz \& de Lapparent 1997). The CFRS (Lilly et al. 1995),  
with $17.5 \leq I_{\mathrm{AB}} \leq 22.5$, extends to $z \simeq 1.3$ 
and this sample includes intrinsically fainter galaxies than the absorber
sample at lower redshifts, $0.20\leq z<0.75$, with
 $-23.5\leq$ \Mab(B)$\leq -17.5$, while at  $0.75\leq z<1.30$ the 
magnitude range covered,  $-23.5\leq$ \Mab(B)$\leq -20.5$, is similar to that
of the high-redshift absorber sample.
The value of \Wr(\OII$\lambda 3727$) averaged per $\sim$0.2 redshift bins
steadily increases with $z$ from $z$=0.35 to 0.9 for both their red and blue
populations  (Le F\`evre et al. 1994) and is roughly equal to ours
at $\left<z\right> \simeq$0.4, but higher by about 30\% than our value for the 
blue subsample at both $\left<z\right> \simeq$ 0.3 and 0.9. This could be a 
consequence of the large statistical uncertainties inherent to  small and/or 
incomplete samples.  An evolution in redshift of the stellar
formation activity is also found by  Songaila et al. (1994) from the analysis
of \Wr(H$\beta$) or \Wr(H$\alpha$+\NII) of a K-band-selected galaxy sample.

\begin{figure}
  \psfig{figure=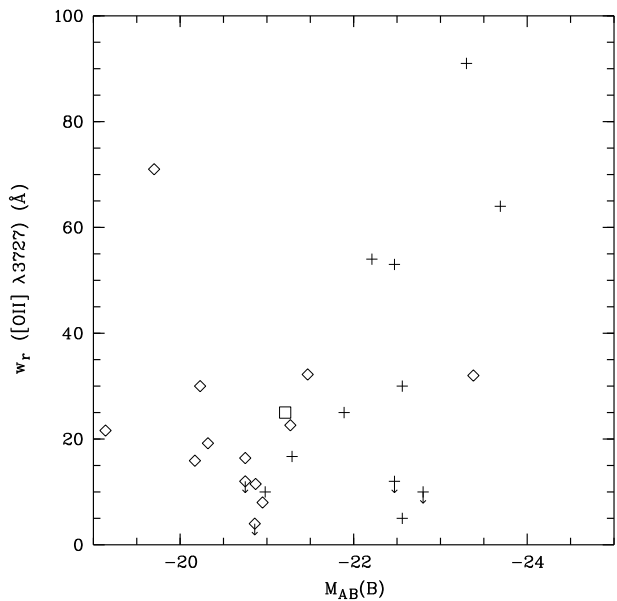,height=8cm}
  \caption[ ]{
    The rest-frame equivalent width of \OII$\lambda 3727$
    versus \Mab(B) and same symbols as in Fig. \ref{FDMb}
    }
  \label{FWMb}
\end{figure}

We have also searched for a possible correlation between
\Wr(\OII$\lambda 3727$) and \Mab(B) (see Fig. \ref{FWMb}). Songaila et al.
(1994) have found that a preferential increase of star formation activity for
the fainter galaxies ($M$(K)$\leq -23.5$) at $0.4 \le z \le 0.8$ is present in
their K-band-selected sample. For our absorber sample, there might be an
opposite trend of increasing \Wr(\OII$\lambda 3727$) for brighter galaxies;
 the null hypothesis that \Wr(\OII$\lambda 3727$) and \Mab(B) are
uncorrelated can be rejected at a level of  91\% for both the Spearman or
Kendall tests. However, excluding only one galaxy from the sample
(G 0334$-$204 at $z$=1.120 with \Wr(\OII$\lambda 3727$)=91\AA\ and
\Mab(B)=$-23.6$) lowers the above significance level to 61\%. For their
larger absorber sample, Steidel et al. (1994) found evidence for a strong
correlation between the rest-frame B$-$K color and $M$(K), fainter galaxies
having on average bleuer colors, but  no such correlation is apparent
 for B$-$K versus $M$(B). They concluded that the difference between the
shape of the B and K luminosity functions of absorbing galaxies is a
consequence of the above correlation, galaxies intrinsically faint in the
infrared being very blue and thus bright optically. This could account for
the lack of correlation between \Wr(\OII$\lambda 3727$) and \Mab(B).

The evolution in redshift of the UV excess blueward of the Balmer discontinuity
as well as \Wr(\OII$\lambda 3727$) for our small sample of
\MgII\ absorption-selected galaxies is consistent with those found for 
larger field galaxy samples which confirms the suggestion that, 
at $z\ge$ 0.2, all luminous field galaxies  have extended gaseous
halos (Bergeron \& Boiss\'e 1991; Steidel et al. 1994).


\begin{acknowledgements}
We would like to deeply thank S. Charlot for fruitful discussions and for having
very kindly made available to us the latest version of the Bruzual \& Charlot
spectral evolution code of stellar populations.
We are also very grateful to the referee, M. Dickinson, for his valuable
comments which have led to substantial improvments to the manuscript.
\end{acknowledgements}


\begin{thebibliography}{}

\bibitem{} Alloin D., Collin-Souffrin S., Joly M., Vigroux L., 1979
  A\&A 78, 200
\bibitem{} Barthel P.D., Tytler D.R., Thomson B., 1990,  A\&AS 82, 339
\bibitem{} Bergeron J., Boiss\'e P., 1991,  A\&A 243, 344
\bibitem{} Bergeron J., Cristiani S., Shaver P., 1992,  A\&A 257, 417
\bibitem{} Bergeron J., Petitjean P., Sargent W.L.W. et al., 1994, ApJ 436, 33
\bibitem[]{} Bergeron J., Stasi\'nska G., 1986, A\&A 169, 1
\bibitem{} Boiss\'e P., Le Brun V., Bergeron J., Deharveng J.-M., 1997,
  in preparation
\bibitem{} Broadhurst T.J., Ellis R.S., Shanks T., 1988,  MNRAS 235, 827
\bibitem{} Bruzual G.A., Charlot S., 1993, ApJ 405, 538
\bibitem{} Bruzual G.A., Charlot S., 1996, in preparation
\bibitem{} Coleman G.D., Wu C., Weedman D.W., 1980,  ApJS 43, 393
\bibitem{} Edmunds M.G., Pagel B.E.J., 1984,  MNRAS 211, 507
\bibitem{} Ellis R.S., 1993,  PASPC 43, 165
\bibitem{} Glazebrook K., Ellis R.S., Colless M.M., Broadhurst T.J.,
  Allington-Smith J.R., Tamvir N.R., Taylor K., 1995, MNRAS 273, 157
\bibitem{} Galaz G., de Lapparent V., 1997, in preparation
\bibitem{} Hunstead R.W., Murdoch H.S., Peterson B.A. et al., 1986,
  ApJ 305, 496
\bibitem{} Le Brun V., Bergeron J., Boiss\'e P., Christian C., 1993,
  A\&A 279, 33
\bibitem{} Le Brun V., Bergeron J., Boiss\'e  P., Deharveng J.-M., 1997,
  A\&A, in press
\bibitem{} Le F\`evre O.,  Lilly S.J., Crampton D., Hammer F., Tresse L.,
  1994 in: B\"ohringer H., Morfill G.E., Tr\"umper J.E. (eds). Seventeenth
  Texas Symposium on Relativistic Astrophysics. New York Academy of Science,
  p. 613
\bibitem{} Lilly S.J., Tresse L., Hammer F., Crampton D., Le F\`evre O.,
  1995, ApJ 455, 108
\bibitem{} Loveday J., Peterson B.A., Efstathiou G., Maddox S.J., 1992,
  ApJ 390, 338
\bibitem{} McGaugh S.S., 1991,  ApJ 380, 140
\bibitem{} McGaugh S.S., 1994,  ApJ 426, 135
\bibitem{} Oke J.B., Gunn J.E., 1983,  ApJ 266, 713
\bibitem{} Pagel B.E.J., Edmunds M.G., Blackwell D.E., Chun M.S.,
  Smith G., 1979,  MNRAS 189,95
\bibitem{} Petitjean P., Bergeron J., , 1990,  A\&A 231, 309
\bibitem{} Sargent  W.L.W., Boksenberg A., Steidel C.C., 1988a,  ApJS 68, 539
\bibitem{} Sargent  W.L.W., Steidel C.C., Boksenberg A., 1988b,  ApJ 334, 22
\bibitem{} Sargent  W.L.W., Steidel C.C., Boksenberg A., 1989,  ApJS 69, 703
\bibitem{} Schade D.J., Lilly S.J., Crampton D., Hammer F.,  Le F\`evre O.,
  Tresse L., 1995, ApJ 451, L1
\bibitem{} Steidel C.C., 1993  in:  Shull J.M., Thronson H.A. (eds). The
  Evolution of Galaxies and their Environment. Kluwer, Dordrecht, p. 263
\bibitem{} Steidel C.C., Bowen D.V., Blades J.C., Dickinson M., 1995,
  ApJ 444, 64
\bibitem{} Steidel C.C., Dickinson M.,  Bowen D.V., 1993,  ApJ 413, L77
\bibitem{} Steidel C.C., Dickinson M., Persson S.E., 1994,  ApJ 437, L75
\bibitem{} Steidel C.C., Sargent W.L.W., 1992, ApJS 80, 1
\bibitem{} Songaila A., Cowie L.L., Hu E.M., Gardner J.P., 1994, ApJS 94, 461
\bibitem{} Ulrich M.-H., 1989,  A\&A 220, 71
\end{thebibliography}
\end{document}